# Multipole-mode solitons in Bessel optical lattices

Yaroslav V. Kartashov,[1] R. Carretero-González,[2] Boris A. Malomed,[3]
Victor A. Vysloukh,[4] and Lluis Torner[1]

[1]ICFO-Institut de Ciencies Fotoniques, Mediterranean Technology Park, 08860 Castelldefels (Barcelona), Spain
[2]Nonlinear Dynamical Systems Group, Department of Mathematics and Statistics, San Diego State University, San Diego, California 92182-7720, USA, http://nlds.sdsu.edu
[3]Department of Interdisciplinary Studies, School of Electrical Engineering, Faculty of Engineering, Tel Aviv University, Tel Aviv 69978, Israel
[4]Departamento de Física y Matemáticas, Universidad de las Americas – Puebla, Puebla 72820, Mexico
vysloukh@mail.udlap.mx

**Abstract:** We study basic properties of quiescent and rotating multipole-mode solitons supported by axially symmetric Bessel lattices in a medium with defocusing cubic nonlinearity. The solitons can be found in different rings of the lattice and are stable when the propagation constant exceeds the critical value, provided that the lattice is deep enough. In a high-power limit the multipole-mode solitons feature a multi-ring structure.

## References and links

1. J. W. Fleischer, T. Carmon, M. Segev, N. K. Efremidis, and D. N. Christodoulides, "Observation of discrete solitons in optically induced real time waveguide arrays," Phys. Rev. Lett. **90**, 023902 (2003).
2. J. W. Fleischer, M. Segev, N. K. Efremidis, and D. N. Christodoulides, "Observation of two-dimensional discrete solitons in optically induced nonlinear photonic lattices," Nature **422**, 147 (2003).
3. D. Neshev, E. Ostrovskaya, Y. Kivshar, and W. Krolikowski, "Spatial solitons in optically induced gratings," Opt. Lett. **28**, 710 (2003).
4. Z. Chen, H. Martin, E. D. Eugenieva, J. Xu, and A. Bezryadina, "Anisotropic enhancement of discrete diffraction and formation of two-dimensional discrete-soliton trains," Phys. Rev. Lett. **92**, 143902 (2004).
5. D. N. Neshev, T. J. Alexander, E. A. Ostrovskaya, Y. S. Kivshar, H. Martin, I. Makasyuk, and Z. Chen, "Observation of discrete vortex solitons in optically induced photonic lattices," Phys. Rev. Lett. **92**, 123903 (2004).
6. J. W. Fleischer, G. Bartal, O. Cohen, O. Manela, M. Segev, J. Hudock, and D. N. Christodoulides, "Observation of vortex-ring discrete solitons in 2D photonic lattices," Phys. Rev. Lett. **92**, 123904 (2004).
7. J. Yang, I. Makasyuk, P. G. Kevrekidis, H. Martin, B. A. Malomed, D. J. Frantzeskakis, and Z. Chen, "Necklacelike solitons in optically induced photonic lattices," Phys. Rev. Lett. **94**, 113902 (2005).
8. B. B. Baizakov, B. A. Malomed, and M. Salerno, "Multidimensional solitons in periodic potentials," Europhys. Lett. **63**, 642 (2003).
9. J. Arlt and K. Dholakia, "Generation of high-order Bessel beams by use of an axicon," Opt. Commun. **177**, 297 (2000).
10. S. H. Tao, X.-C. Yuan, and B. S. Ahluwalia, "The generation of an array of nondiffracting beams by a single composite computer generated hologram," J. Opt. A: Pure. Appl. Opt. **7**, 40 (2005).
11. Y. V. Kartashov, V. A. Vysloukh, and L. Torner, "Rotary solitons in Bessel optical lattices," Phys. Rev. Lett. **93**, 093904 (2004).
12. Y. V. Kartashov, V. A. Vysloukh, and L. Torner, "Stable ring-profile vortex solitons in Bessel optical lattices," Phys. Rev. Lett. **94**, 043902 (2005).
13. Y. V. Kartashov, A. A. Egorov, V. A. Vysloukh, and L. Torner, "Rotary dipole-mode solitons in Bessel optical lattices," J. Opt. B: Quantum Semiclass. Opt. **6**, 444 (2004).
14. Y. V. Kartashov, A. A. Egorov, V. A. Vysloukh, and L. Torner, "Stable soliton complexes and azimuthal switching in modulated Bessel optical lattices," Phys. Rev. E **70**, 065602(R) (2004).
15. D. Mihalache, D. Mazilu, F. Lederer, B. A. Malomed, Y. V. Kartashov, L.-C. Crasovan, and L. Torner, "Stable spatiotemporal solitons in Bessel optical lattices," Phys. Rev. Lett. **95**, 023902 (2005).
16. A. Desyatnikov, D. Neshev, E. Ostrovskaya, Y. S. Kivshar, W. Krolikowski, B. Luther-Davies, J. J. Garcia-Ripoll, and V. Perez-Garcia, "Multipole spatial vector solitons," Opt. Lett. **26**, 435 (2001).
17. A. S. Desyatnikov, D. Neshev, E. A. Osrovskaya, Y. S. Kivshar, G. McCarthy, W. Krolikowski, and B. Luther-Davies, J. Opt. Soc. Am. B **19**, 586 (2002).

18. A. M. Glass and J. Strait, in "Photorefractive Materials and their Applications I," P. Günter and J. P. Huignard eds., (Springer, Berlin, 1988) pp. 237-262.

Two-dimensional (2D) spatial optical solitons of various types, supported by periodic lattices in nonlinear media, have recently attracted a great deal of attention. This was stimulated, first of all, by the possibility to create such solitons in photorefractive crystals, where an effective photonic lattice is induced by the linear-superposition pattern generated by laser beams illuminating the crystal in the ordinary polarization, while soliton is observed in the extraordinary polarization [1-7]. Optical lattice is also an important tool to study nonlinear pattern formation in Bose-Einstein condensates (BECs), where it can support multidimensional solitons, stabilizing them against collapse in self-attracting BECs [8].

Another possibility for the creation of 2D solitons is offered by a concentric lattice with the radial distribution of the intensity obeying the pattern of the Bessel function. Such a cylindrical lattice can be created by nondiffracting linear optical beams [9,10]. Several types of solitons in this setting have been studied theoretically, assuming focusing or defocusing nonlinearity of the medium [11-15]. In particular, *rotary solitons*, i.e., strongly localized wave packets rotating at a constant angular velocity in the Bessel lattice, are possible in the latter case [11], and stable vorticity-carrying ring-shaped (azimuthally uniform) solitons were predicted in the case of defocusing [12]. *Dipole-mode solitons* that may be regarded as bound states of two rotary ones, set at diametrically opposite positions in the same circular trough, were also predicted for the focusing nonlinearity [13].

A nontrivial possibility that has not been explored before and is the subject of the present work, is a possibility to construct stable multipole-mode solitons, featuring an azimuthal structure, in the system combining the cubic *defocusing* nonlinearity and a Bessel lattice (we stress once again that all the states in axially symmetric defocusing media considered thus far were uniform in the azimuthal direction). We aim to construct such patterns of two most fundamental types, viz., dipole- and quadrupole-mode solitons, both static and rotating, and investigate their stability.

The concept of multi-pole solitons in two dimensions was first explored in Refs [16,17], where stabilization of complex multi-pole structures was achieved due to XPM (cross-phase-modulation) coupling with a nodeless component in a vectorial (two-component) system. However, such solitons require *a focusing* saturable nonlinearity for their existence.

Following Refs. [11-15], we assume that an optical beam propagates along the $\xi$ axis in a bulk medium with defocusing cubic nonlinearity and imprinted transverse modulation of refractive index. This setting is described by the nonlinear Schrödinger equation for the complex field amplitude $q$:

$$i\frac{\partial q}{\partial \xi} = -\frac{1}{2}\left(\frac{\partial^2 q}{\partial \eta^2} + \frac{\partial^2 q}{\partial \zeta^2}\right) + q|q|^2 - pR(\eta,\zeta)q. \qquad (1)$$

In Eq. (1), the longitudinal $\xi$ and transverse $\eta,\zeta$ coordinates are scaled to the diffraction length and the width of the input beam, respectively, and $p$ measures the depth of the optical lattice. We suppose that optical lattice is created by a first-order Bessel beam, whose field is given by $J_1[(2b_{\text{lin}})^{1/2}r]\exp(-ib_{\text{lin}}\xi + i\phi)$, where $r^2 = \eta^2 + \zeta^2$, $\phi$ is the azimuthal angle, and $b_{\text{lin}}$ determines the transverse scale of the lattice. Accordingly the lattice profile is determined by the beam intensity in the form of $R(\eta,\zeta) = J_1^2[(2b_{\text{lin}})^{1/2}r]$.

Nondiffracting Bessel beams of different orders can be created experimentally in a number of ways, including illumination through a narrow annular slit placed in the focal plane of a lens or axicon, or by dint of holographic techniques [9,10]. The interaction of beams with the orthogonal polarizations in a nonlinear medium can be utilized for guiding spatial solitons in the concentric lattice. In particular, a Bessel photonic lattice can be induced by an ordinarily polarized nondiffracting beam in a photorefractive medium, while the soliton propagates in the extraordinary polarization. Because of the strong anisotropy of the nonlinear response, the Bessel beam does not feel the material nonlinearity and thus remains undistorted, while the extraordinarily polarized probe beam experiences strong nonlinearity.

Notice that Eq. (1) implies opposite signs of XPM and SPM (self-phase-modulation) nonlinearities, which occurs in semiconductor photorefractive crystals, such as GaAs:Cr, InP:Fe, and CdTe:In, that belong to the $\bar{4}3m$ point symmetry group [18]. These materials are transparent for near-infrared wavelengths, and exhibit strong photorefractivity ($n^3 r_{41} = 152$ pm/V in CdTe:In), while nonlinearity sign may be changed by a $\pi/2$-rotation of the polarization direction. In such crystals, biased by a sufficiently strong static electric field $E_0 \sim 10^5$ V/m, and for laser beams with the width $\sim 10$ $\mu$m, the propagation distance $\xi = 1$ corresponds to about 1 mm in physical units, while the dimensionless amplitude $q \sim 1$ corresponds to peak intensities $\sim 50$ mW/cm$^2$.

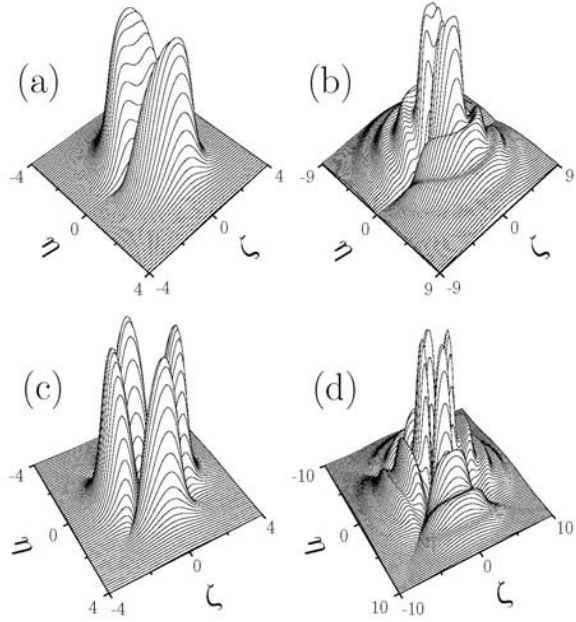

Fig. 1. Profiles of dipole-mode solitons at $b = 2.49$, $p = 15$ (a), and $b = 0.42$, $p = 15$ (b). Profiles of quadrupole-mode solitons at $b = 2.79$, $p = 20$ (c), and $b = 0.58$, $p = 20$ (d).

Besides the implementation in optics, Eq. (1) with the same Bessel potential may also be interpreted as Gross-Pitaevskii equation for an effectively 2D BEC trapped in a optical lattice created by the linear Bessel beam. In terms of the BEC, the defocusing corresponds to positive scattering length of atomic collisions (repulsion) in the BEC, while the necessary sign of the effective potential in Eq. (1) is provided by choosing the carrier frequency of the optical beam to be red-detuned relative to the frequency of the dipole transition in the condensate atoms. In either case (optics or BEC), the soliton-carrying field produces no feedback on Bessel lattice. Equation (1) conserves the total power (energy flow, or number of atoms in the BEC):

$$U = \int\limits_{-\infty}^{\infty} \int\limits_{-\infty}^{\infty} |q|^2 \, d\eta d\zeta. \tag{2}$$

We search for multipole-mode solutions of Eq. (1) numerically (using the standard relaxation method) in the form of $q(\eta,\zeta,\xi) = w(\eta,\zeta)\exp(ib\xi)$, where $b$ is the propagation constant ($-b$ is the chemical potential in the BEC), and $w(\eta,\zeta)$ is real. Further, without loss of generality, we fix the transverse lattice scale by setting $b_{\text{lin}} = 1$ (this can be always done by obvious rescaling of the transverse coordinates in Eq. (1)), and vary $b$ and $p$.

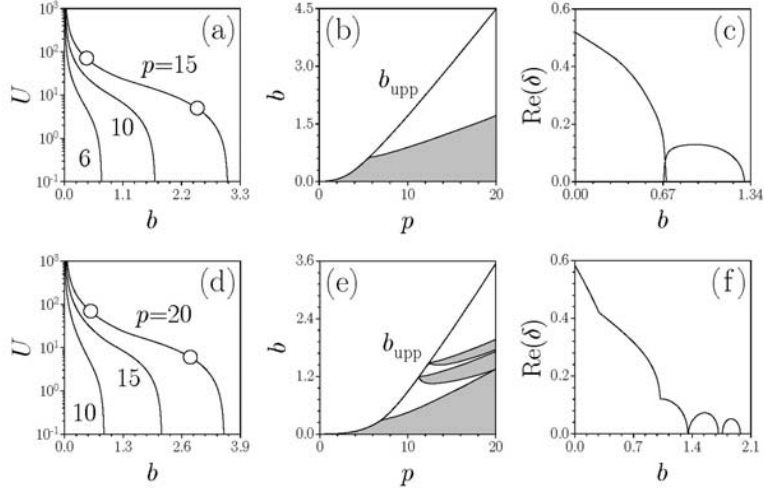

Fig. 2. (a) Energy flow vs propagation constant for the dipole-mode solitons. (b) Stability and instability (shaded) domains in the $(p,b)$ plane. (c) The real part of the perturbation growth rate vs propagation constant at $p = 15$. (d) Energy flow vs the propagation constant and (e) stability and instability domains in the $(p,b)$ plane for the quadrupole-mode solitons. (f) The real part of the perturbation growth rate vs the propagation constant at $p = 20$. Points marked by circles in (a) and (d) correspond to the profiles shown in Fig. 1.

To examine the linear soliton stability, we then take perturbed solutions to Eq. (1) as $q = (w+\varepsilon[u\exp(\delta\xi)+v\exp(\delta^*\xi)])\exp(ib\xi)$, with an infinitesimal amplitude $\varepsilon$, perturbation eigenmodes $u$ and $v$, and instability growth rate $\delta$ (* stands for the complex conjugation). Substitution of this expression in Eq. (1) and linearization around the stationary solution leads to equations

$$\begin{aligned}
i\delta u - bu + \frac{1}{2}\left(\frac{\partial^2 u}{\partial \eta^2} + \frac{\partial^2 u}{\partial \zeta^2}\right) - w^2 v^* - 2w^2 u + pRu = 0, \\
i\delta v - bv + \frac{1}{2}\left(\frac{\partial^2 v}{\partial \eta^2} + \frac{\partial^2 v}{\partial \zeta^2}\right) - w^2 u^* - 2w^2 v + pRv = 0,
\end{aligned} \tag{3}$$

from which the eigenvalues $\delta$ were obtained numerically.

We have found a variety of multipole-mode solutions of Eq. (1). In most cases, they reside in the first ring of the first-order Bessel lattice. Figure 1 displays, in terms of $|w(\eta,\zeta)|$, typical profiles of dipole- and quadrupole-mode solitons, which seem, respectively, as patterns built of two or four localized spots, with opposite signs of the field in adjacent spots (hence the spots repel each other, which helps to stabilize the patterns). Despite the defocusing character of the nonlinearity, solitons are well localized in the radial direction because of the confining action of the lattice. On the other hand, the multipole-mode solitons are not tightly localized in the azimuthal direction, in contrast to the dipole-mode solitons in the Bessel lattice with the focusing nonlinearity [13].

In the low-amplitude limit, the multipole-mode solitons go over into linear modes guided by the first-order Bessel lattice (Figs. 1(a) and 1(c)). Note that, with decrease of the lattice depth, the corresponding linear modes become broader and may occupy several rings of the lattice. In the large-amplitude limit, solitons always expand over several rings of the lattice, see Figs. 1(b) and 1(d). With the growth of the total power, their amplitude only slightly increases, while expansion in radial direction is conspicuous. This behavior finds its manifestation in a monotonic growth of the power with the decrease of the propagation constant (Figs. 2(a) and 2(d)). There are lower and upper borders (cutoffs) for the existence of the multipole-mode solitons. The power diverges and vanishes as $b$ approaches, respectively, the lower and upper cutoffs, $b_{\text{low}} \equiv 0$ and $b_{\text{upp}}$. It was found that $b_{\text{upp}}$ is a monotonically growing function of the lattice's depth $p$, see Figs. 2(b) and 2(e). At a fixed lattice depth, the dipole-mode solitons have a larger existence interval (in terms of the propagation constant $b$) than the quadrupole-mode solitons and higher-order ones. In fact, the existence domain gradually shrinks with the increase of the soliton order, compare Figs. 2(b) and 2(e). Besides solitons residing in the first ring of the Bessel lattice (the one closest to the center)e, we have also found families of the dipole- and quadrupole-mode solitons, including stable ones, entrenched father from the center, in the rings of a larger radius.

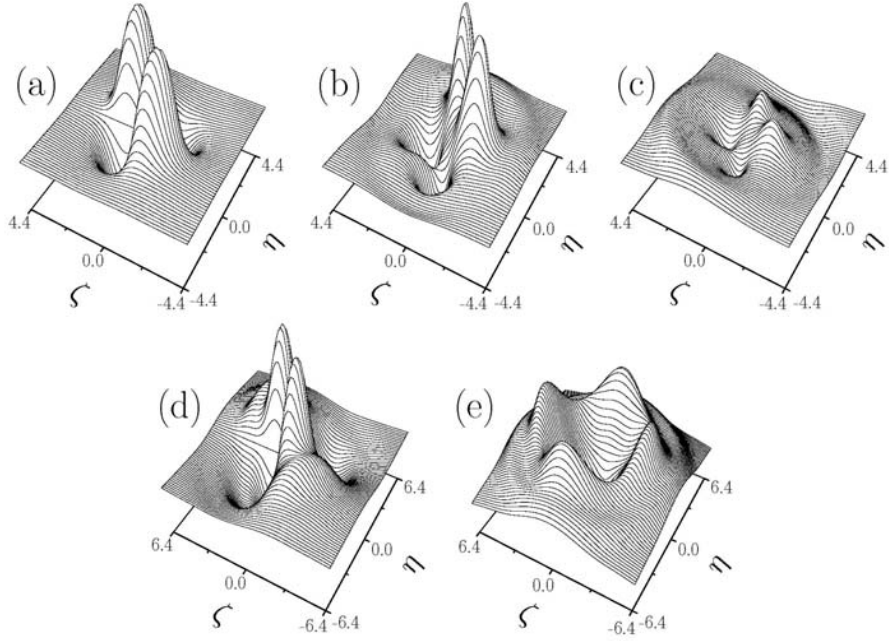

Fig. 3. (a) Profile of quadrupole-mode soliton at $b = 1.9$, $p = 20$. Real (b) and imaginary (c) parts of perturbation associated with oscillatory instability of soliton depicted in (a) with $\delta = 0.048 + 0.721i$. (d) Profile of quadrupole-mode soliton at $b = 0.4$, $p = 10$. (e) Real part of perturbation associated with exponential instability of soliton depicted in (d) with $\delta = 0.043$. Panels (b) and (c) are shown with the same vertical scale.

A comprehensive linear stability analysis has revealed that the dipole-mode solitons are completely stable if the propagation constant exceeds a certain critical value $b_{\rm cr}$, provided that the lattice is deep enough ($p > 5.7$). The remaining instability domain for the dipoles is shaded in Fig. 2(b). Its upper boundary slowly grows with the lattice depth, so that the width of the stability domain $b_{\rm upp} - b_{\rm cr}$ also increases with $p$. Unstable dipole-mode solitons are subject to both monotonously growing and oscillatory instabilities. Interestingly, there exist parameter regions where these two types of instability coexist. For example, in Fig. 2(c), the left domain ($b < 0.67$) features the monotonous instability, while the right domain corresponds to its oscillatory counterpart, with a very narrow region of coexistence between them. In contrast, for a shallower lattice, the oscillatory instability domain extends up to $b_{\rm low}$ and strongly overlaps with the domain of the monotonous instability. For both types of the instability, $\text{Re}(\delta) \to 0$ as $b \to b_{\rm cr}$. Note, that in the case of weak instability, a characteristic length of the instability development $\xi \sim 1/\text{Re}(\delta)$ may be much larger than actual length available to the experiment, hence weakly unstable soliton may also be observed.

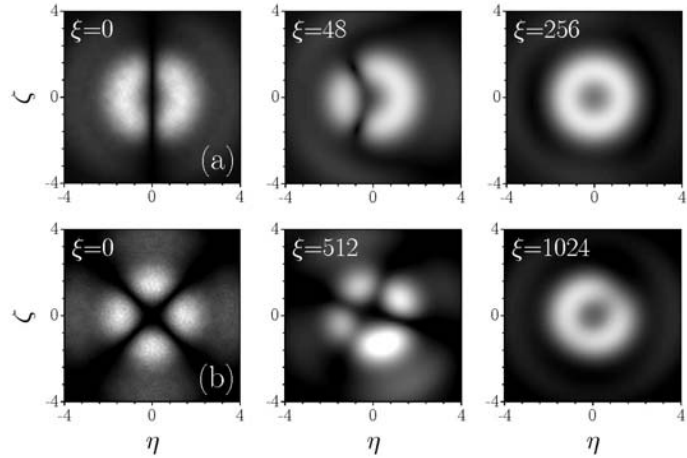

Fig. 4. Evolution of an unstable (a) dipole-mode soliton for $b = 0.9$, $p = 15$, and (b) its quadrupole-mode counterpart for $b = 0.4$, $p = 10$. In both cases, white noise with variance $\sigma_{\rm noise}^2 = 0.01$ was added as initial perturbation.

The $(p, b)$ stability diagram for the quadrupole-mode solitons is more complex, see Fig. 2(e). Besides the main stability domain that is found at $p \approx 12.4$, there are several narrow stability windows that gradually merge with increase of the lattice depth $p$. The corresponding dependence $\delta(b)$ is quite complicated, especially for deep lattices, see Fig. 2(f). In Fig. 2(f), the real growth rates dominate for $b \to b_{\rm low}$, while the oscillatory instability takes place closer to the upper cutoff. Similar stability properties were found for higher-order solitons. Profiles of instability eigenmodes for the quadrupole-mode solitons are displayed in Fig. 3, which shows that, for all instability types, the perturbation profiles feature local extrema at spots where the soliton intensity is smallest.

To verify predictions of the linear stability analysis, we directly solved Eq. (1) with an initial perturbation, $q(\eta, \zeta, \xi = 0) = w(\eta, \zeta)[1 + \rho(\eta, \zeta)]$, where $w(\eta, \zeta)$ corresponds to the stationary solution, and $\rho(\eta, \zeta)$ stands for broadband noise with the Gaussian distribution and variance $\sigma_{\rm noise}^2$. In all the cases studied, the predictions of the linear stability analysis were confirmed. It was found that unstable multipole-mode solitons gradually loose their symmetry and transform into a ground-state (azimuthally uniform) soliton sitting in the first ring of the

Bessel lattice (cf. Fig. 4), that was studied in Ref. [12]. In contrast, stable solitons retain their structure indefinitely long, even in the presence of strong input noise (Fig. 5). Moreover, we have also found that the stable solitons survive in the presence of perturbations that cause strong bending of the dark stripes in the solitons' structure. Depending on the stripe deformation, the perturbed soliton quickly restores its unperturbed profile or features long-lived vibrations that do not destroy its topological structure.

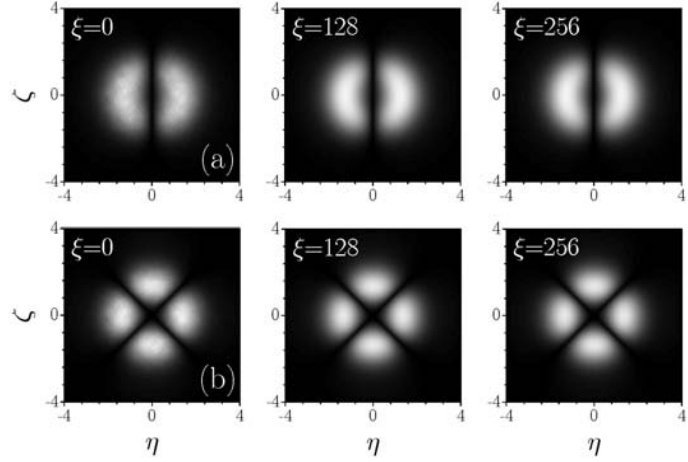

Fig. 5. Stable propagation of a dipole-mode soliton at $b = 2$, $p = 15$ (a) and quadrupole-mode one at $b = 2$, $p = 20$ (b). In both cases, white noise with variance $\sigma_{\text{noise}}^2 = 0.01$ was added initially.

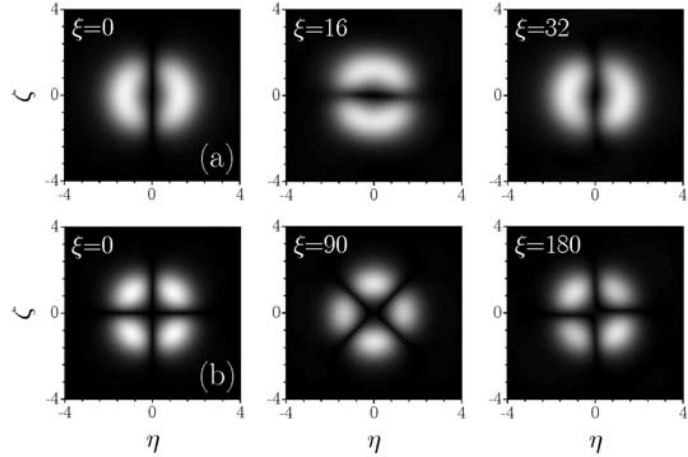

Fig. 6. Stable clockwise rotation of a dipole-mode soliton corresponding to $b = 2$, $p = 15$ (a), and quadrupole-mode one corresponding to $b = 2.79$, $p = 20$ (b).

Multipole-mode solitons can be set in rotation, by imprinting a phase twist onto the input field distribution, or by pushing out-of-phase fundamental beams, which form the soliton, in opposite directions in $(\eta, \zeta)$ plane. In this case, a steadily rotating pattern emerges, that survi-ves indefinitely many rotations, without any tangible loss (Fig. 6). Rotary states of such multi-spot patterns suggest new possibilities for all-optical routing of light beams.

Summarizing, we have found that cylindrical optical lattices induced by nondiffracting Bessel beams in medium with defocusing cubic nonlinearity support various stable multipole-mode solitons. The existence of such complex patterns, that can also stably rotate, essentially extends the variety of phenomena generated by Bessel optical lattices. The results suggest new ways of manipulating light signals, and the creation of novel localized states in BECs.

**Acknowledgements**

This work was partially supported by the Government of Spain through grant BFM 2002-2861 and by the Ramon-y-Cajal program. RCG and BAM appreciate hospitality of ICFO. RCG acknowledges the support from NSF-DMS-0505663 and a SDSU-GIA grant provided by the SDSU Research Foundation. VAV recognizes support from CONACYT grant 46552. The work of BAM was partly supported by the Israel Science Foundation through the Excellence-Center grant No. 8006/03.